\documentclass[twocolumn]{aastex63}

\shorttitle{Missing Titanium in W49B}
\shortauthors{Sato et al.}
\graphicspath{{./}{figures/}}

\begin{document}

\title{Missing Titanium in the Asymmetric Supernova Remnant W49B}

\correspondingauthor{Toshiki Sato}
\email{toshiki@meiji.ac.jp}

\author[0000-0001-9267-1693]{Toshiki Sato}
\affil{Department of Physics, School of Science and Technology, Meiji University, 1-1-1 Higashi Mita, Tama-ku, Kawasaki, Kanagawa 214-8571, Japan}

\author[0000-0003-2008-6887]{Makoto Sawada}
\affiliation{Department of Physics, Rikkyo University, 3-34-1 Nishi-Ikebukuro, Toshima-ku Tokyo 171-8501, Japan}

\author[0000-0003-2611-7269]{Keiichi Maeda}
\affiliation{Department of Astronomy, Kyoto University, Kitashirakawa-Oiwake-cho, Sakyo-ku, Kyoto 606-8502 Japan}

\author[0000-0002-8816-6800]{John P.Hughes}
\affiliation{Department of Physics and Astronomy, 
Rutgers University, 136 Frelinghuysen Road, Piscataway, 
NJ 08854-8019, USA}

\author[0000-0003-2063-381X]{Brian J. Williams}
\affiliation{NASA, Goddard Space Flight Center, 8800 Greenbelt Road, Greenbelt, MD 20771, USA}



\begin{abstract}

The progenitor of the W49B supernova remnant is still under debate. One of the candidates is a jet-driven core-collapse supernova. In such a highly asymmetric explosion, a strong $\alpha$-rich freezeout is expected in local high entropy regions, which should enrich elements synthesized by the capture of $\alpha$-particles such as $^{44}$Ti and $^{48}$Cr (decaying to $^{44}$Ca and $^{48}$Ti, respectively). In the present work, in order to infer the progenitor of the W49B remnant, we constrain the amount of stable Ti ($^{48}$Ti) synthesized, using the {\it Suzaku} observation. We found no firm evidence for the Ti line and set the upper limit of $M_{\rm Ti}/M_{\rm Fe} < 8.2 \times$ 10$^{-4}$ (99\% limit using Xspec) and $M_{\rm Ti}/M_{\rm Fe} < 1.9 \times$ 10$^{-3}$ (99\% limit using SPEX), and thus excluded almost all hypernova/jet-driven supernova models. Our results, as complemented by some previous studies, suggest that a Type Ia supernova from a near-$M_{\rm Ch}$ (Chandrasekhar mass) white dwarf is the most favorable candidate for the origin of W49B. Future observations with X-ray calorimeter missions, such as XRISM, will give us a stronger constraint on the progenitor. 

\end{abstract}


\section{Introduction} \label{sec:intro}
W49B is a peculiar supernova remnant (SNR) that exhibits highly asymmetric ejecta distribution \citep[e.g.,][]{2009ApJ...691..875L,2011ApJ...732..114L,2013ApJ...764...50L} and recombining plasma \citep[e.g.,][]{2009ApJ...706L..71O,2010A&A...514L...2M,2018ApJ...868L..35Y,2020ApJ...893...90S,2020ApJ...903..108H}. While the progenitor of the remnant has been a focus of much attention, a firm conclusion has not yet been reached. The elongated structure of Fe-rich ejecta has been suggested to be related to a bipolar/jet-driven Type Ib/Ic explosion and/or interactions between the shock and a surrounding interstellar cloud \citep[][]{2007ApJ...654..938K,2013ApJ...764...50L}. On the other hand, X-ray spectral studies have argued for the Type Ia supernova origin of W49B from its element abundances \citep[][]{2000ApJ...532..970H,2018A&A...615A.150Z,2020ApJ...904..175S}. The main difficulty in inferring the progenitor comes from the lack of emissions below 1 keV due to strong interstellar absorption ($N_{\rm H} \sim 5\times10^{22}~{\rm cm}^{-2}$); without information from O-Ne-Mg-rich ejecta seen in soft X-rays that can characterize a massive progenitor for a core-collapse SN, it is difficult to distinguish from one progenitor scenario to another, since the abundance pattern from Si to Fe is strongly dependent on explosive nucleosynthesis. 

W49B is also known as the remnant with the brightest Fe-K$\alpha$ line in the Galaxy \citep{2014ApJ...785L..27Y}, providing a unique opportunity to test the explosive nucleosynthesis around Fe-group elements. The Fe-group elements are mainly synthesized by Si burning around the core of exploding stars. 
This allows us to investigate the differences in the central environment of core-collapse supernovae and Type Ia supernovae \citep[e.g.,][]{1973ApJS...26..231W}. In particular, asymmetric/energetic explosions of massive stars are believed to produce a large amount of $^{44}$Ti, since the stronger $\alpha$-rich freeze-out could occur in the region where the larger energy is deposited \citep{1998ApJ...492L..45N,2001ApJ...555..880N,2003ApJ...598.1163M}. 

The $\alpha$-rich freeze-out is a nuclear burning regime that describes explosive nucleosynthesis in a high-entropy region (especially for core-collapse supernovae). In this regime, the abundant $\alpha$ particles ($^{4}$He) are captured by heavy elements, and heavier elements are synthesized. Thus, the coexistence of the main product Fe ($^{56}$Fe after decay of $^{56}$Ni) and other $\alpha$ elements (e.g., $^{44}$Ti) in the ejecta can be used as strong evidence for the high-entropy process.
Not only the most famous $\alpha$-rich freeze-out product, radioactive $^{44}$Ti \citep[e.g.,][]{2014Natur.506..339G}, some rare stable elements synthesized through captures of $\alpha$-particles, such as $^{48}$Ti (after decay of $^{48}$Cr), can also characterize the high-entropy nuclear burning regime \citep{2001ApJ...555..880N,2018ApJ...852...40W,2021Natur.592..537S,2022PASJ...74..334I}. The detection of these elements synthesized in the high-entropy nuclear burning regime can provide strong evidence of highly asymmetric explosions.

In this paper, we search for the Ti-K$\alpha$ line in the X-ray spectrum of W49B. If the remnant originated from an asymmetric core-collapse supernova, we expect to see the Ti-K line in the spectrum as observed in Cassiopeia A \citep{2021Natur.592..537S,2022PASJ...74..334I}. Even if the origin is the Type Ia supernova, estimating the amount of Ti could provide useful information about the progenitor white dwarf (WD) \citep[e.g.,][]{2021ApJ...913L..34O}. While Mn has mainly been used to characterise Type Ia progenitors, the argument can be complicated since Mn production is sensitive to the neutrino process in core-collapse supernovae \citep{2023ApJ...954..112S}. Our new approach will provide another independent tool to distinguish between Type Ia and core-collapse progenitors.

\section{Upper Limit of Taitanium in W49B} \label{sec:obs}

\begin{figure}[t]
 \begin{center}
  \includegraphics[bb=0 0 944 958, width=8cm]{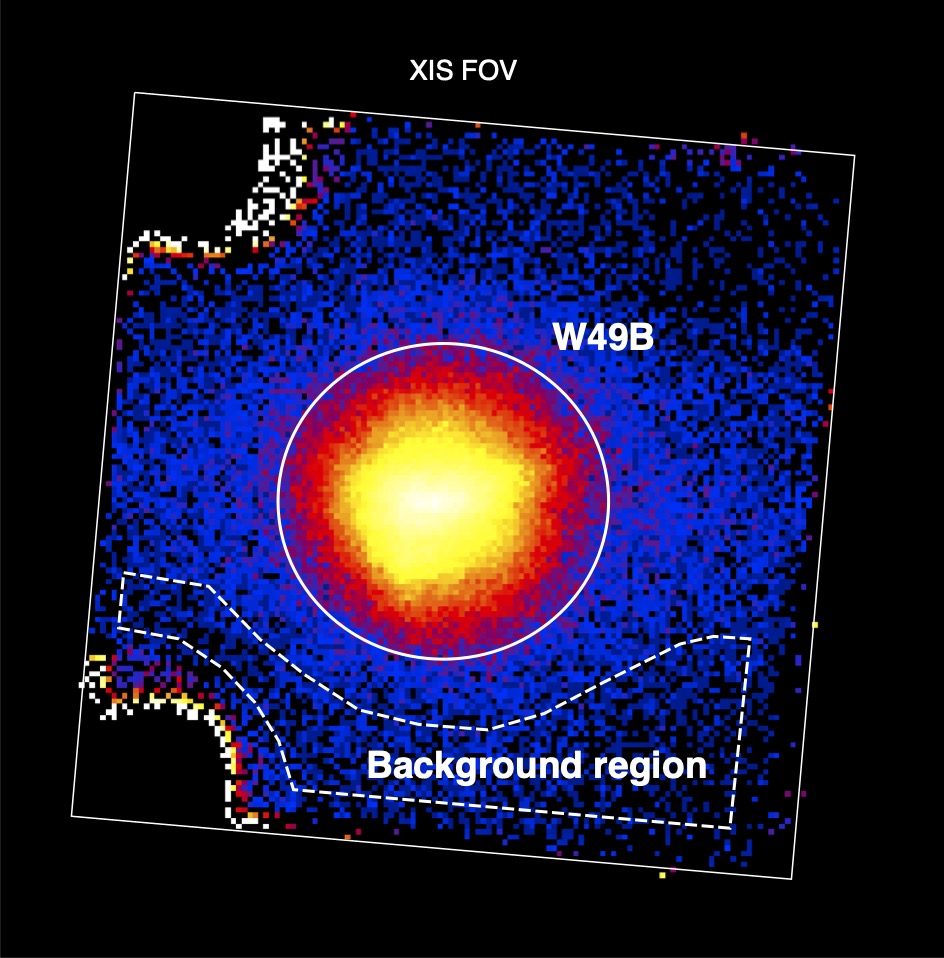}
 \end{center}
\caption{X-ray image of W49B taken with {\it Suzaku}. The white circle was used as the source region for the spectral analysis. The dashed area shows the background region.}
\label{fig:w49b}
\end{figure}

\begin{figure*}[t!]
 \begin{center}
  \includegraphics[bb=0 0 1488 598, width=16cm]{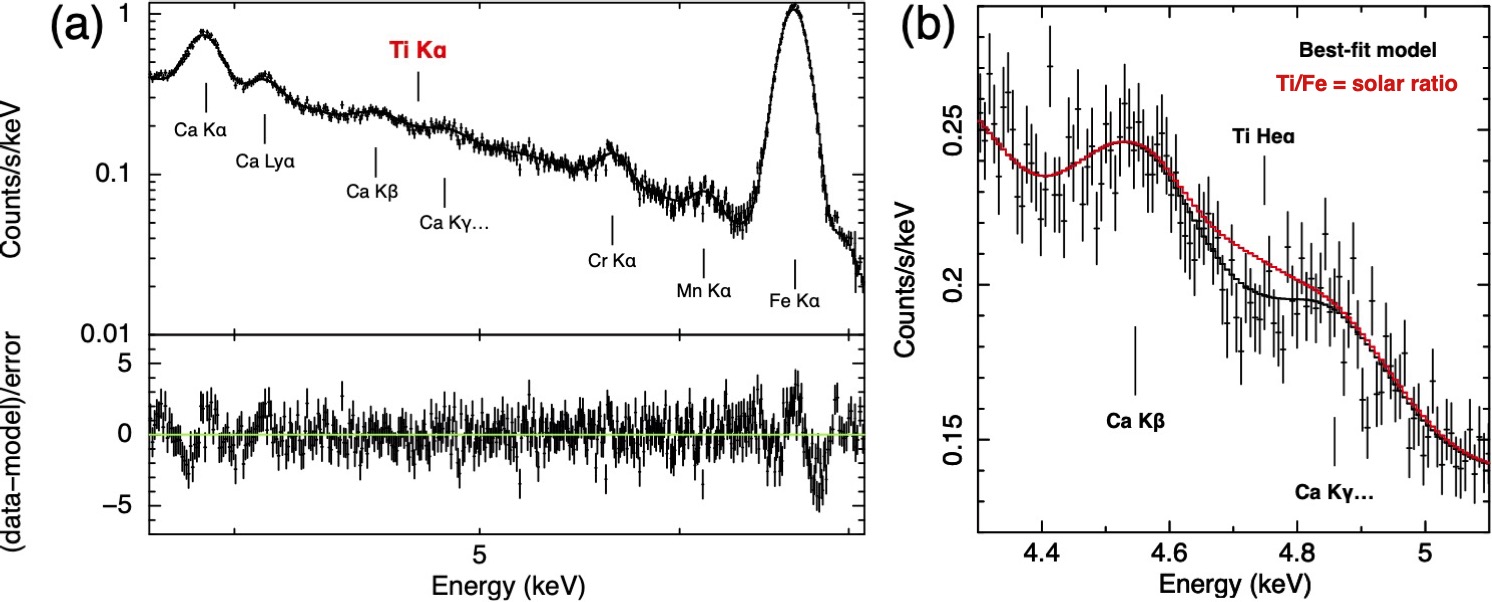}
 \end{center}
\caption{(a) X-ray spectrum of W49B in 3.7--7.1 keV. The solid curve shows the best-fit model. The bottom window shows the residuals of the fitting. (b) The zoom-up view around the Ti-K$\alpha$ line. The black curve shows the best-fit model. If we assume the solar value of Ti/Fe, the model deviates significantly from the data.}
\label{fig:f1}
\end{figure*}

W49B has been observed by {\it Suzaku} in 2009. We only use data from the front-illuminated CCDs \citep[XIS0 and XIS3;][]{2007PASJ...59S..23K}. The data were reprocessed with the recent calibration database (2016 February 14). We applied the standard screening criteria to create the cleaned event list. The total exposure time is 114 ks. 
Figure \ref{fig:w49b} shows the X-ray image of W49B taken with {\it Suzaku}. We extracted the X-ray spectrum (Figure \ref{fig:f1} (a)) from the ellipse region shown in the figure. The background spectrum was chosen from the blank sky around the remnant (see the broken line in Figure \ref{fig:w49b}).

We fitted the spectrum with an absorbed vvrnei model in XSPEC 12.11.0 (atomdb version 3.0.9), which reproduced the observation well ($\chi^2$/d.o.f $= 1.59$). Here, the initial electron temperature is fixed at 4 keV as estimated in \cite{2018ApJ...868L..35Y}. The fitting parameters are summarized in Table \ref{tab1}, where the values are similar to those in the previous studies \citep[e.g.,][]{2018A&A...615A.150Z}.
We found that there is no significant Ti emission and set the upper limit of $M_{\rm Ti}/M_{\rm Fe} < 8.2 \times$ 10$^{-4}$ (99\% confidence level). While the Ti-K$\alpha$ line is expected to lie between the Ca-K$\beta$ and Ca-K$\gamma$ lines, there is a clear valley structure between them (Figure \ref{fig:f1}(b)). In addition, using SPEX, we could not find a firm detection of the Ti line in the X-ray spectrum of W49B (the best-fit abundance of Ti is still zero), either. In this case, we set 99\% and 90\% upper limits of Ti/Fe $<$ 1.9$\times$10$^{-3}$ and $<$ 1.5$\times$10$^{-3}$, respectively. The upper limits are slightly higher than with xspec, which would result from some different fitting situations, such as different line emissivities, different reproducibility of the continuum emission, and so on. The fitting was well ($\chi^2$/d.o.f $= 1.62$), although it is a little worse than that with xspec. In the following sections we mainly use the results with Xspec because of the better fit. In either case, there is no significant detection of Ti. This suggests that the SN explosion for W49B was not an explosion that efficiently supplied Ti. The detailed discussions are given in the following sections.

\begin{table}[t]
\caption{The best-fit parameters for the iron-rich ejecta in Fig.\ref{fig:f1}. The errors show 99\% confidence level ($\Delta\chi^2$ = 6.63). The solar abundance in \cite{1989GeCoA..53..197A} is used.}
\begin{center}
\begin{tabular}{lr}
\hline
$n_{\rm H}$ (fixed) [cm$^{-2}$]                 &   5$\times$10$^{22}$     \\
$kT_{\rm e}$ [keV]         &   1.06$\pm$0.03  \\
Initial $kT_{\rm e}$ [keV] (fixed)       &  4  \\
${\rm [Ar/H]/[Ar/H]}_{\odot}$                   &   5.0$^{+0.4}_{-0.5}$\\
${\rm [Ca/H]/[Ca/H]}_{\odot}$                   &   4.7$\pm$0.2\\
${\rm [Ti/H]/[Ti/H]}_{\odot}$                   &   $<$ 1.82\\
${\rm [Cr/H]/[Cr/H]}_{\odot}$                   &   6.8$\pm$1.0\\
${\rm [Mn/H]/[Mn/H]}_{\odot}$                   &   11.7$\pm$2.3\\
${\rm [Fe/H]/[Fe/H]}_{\odot}$                   &   4.0$^{+0.4}_{-0.2}$\\
$n_{e}t$ [cm$^{-3}$ s]               &   3.1$^{\pm}$0.2 $\times$10$^{11}$\\
norm [$\frac{10^{-14}}{4 \pi D^2} \int n_{e} n_{\rm H} dV$ cm$^{-5}$] & (7.5$^{+0.6}_{-0.4}$)$\times$10$^{-1}$\\
$\chi^2$/d.o.f.                                 & 716.56/450 \\
\hline
\end{tabular}
\label{tab1}
\end{center}
\end{table}

\begin{figure*}[t!]
 \begin{center}
  \includegraphics[bb=0 0 1500 1010, width=16cm]{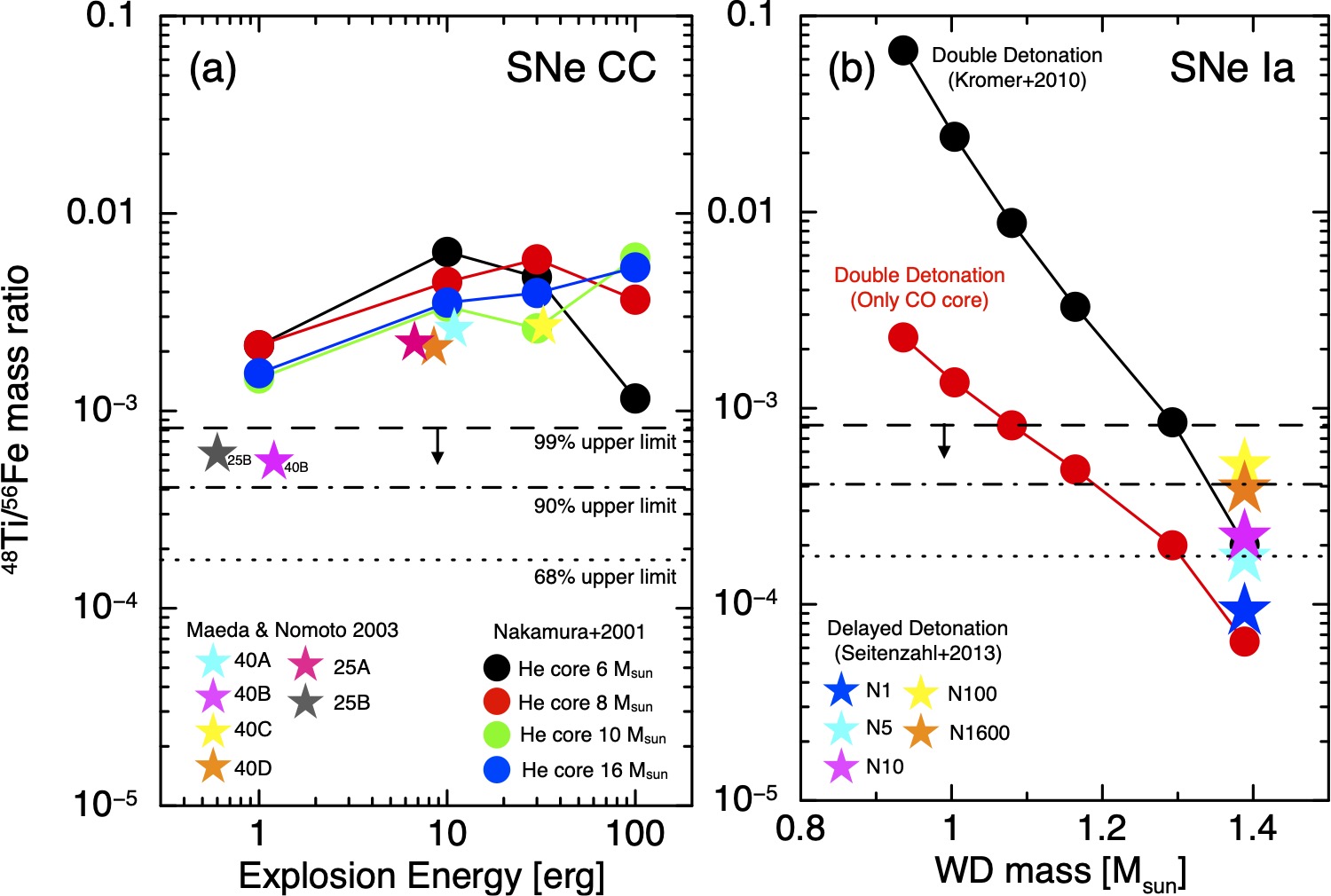}
 \end{center}
\caption{The observed Ti/Fe mass ratio comparing with those of theoretical models. The dashed, dash-dotted and dotted lines show 99\%, 90\%, and 68\% upper limits, respectively. (a) The theoretical calculations of the Ti/Fe mass ratio for core-collapse supernovae in \cite{2003ApJ...598.1163M} and \cite{2001ApJ...555..880N}, where the horizontal axis shows the explosion energy. (b) The theoretical calculations of the Ti/Fe mass ratio for Type Ia supernovae in \cite{2010ApJ...719.1067K} and \cite{2013MNRAS.429.1156S}, where the horizontal axis shows the mass of the WD progenitor.}
\label{fig:f2}
\end{figure*}




\section{Comparison with theoretical models} \label{sec:dis}
In this study, we found that there is no clear signature of stable Ti in the X-ray spectrum of W49B observed with {\it Suzaku}. We here set the tight upper limit of $M_{\rm Ti}/M_{\rm Fe} < 8.2 \times$ 10$^{-4}$, which is significantly lower than the solar value. This means that the type of the SN explosion responsible for the W49 remnant does not form a main source of Ti in the Galaxy. Here we discuss the origin of W49B based on the comparisons with theoretical models.

Ti is produced at the deepest parts (Si burning) of either Type Ia\footnote{In the case of the double-detonation models, Ti is also produced at the WD surface by He burning.} and core-collapse supernovae. Therefore, to trace the signature of the Ti production through observations, the innermost ejecta must already be heated by the reverse shock. The age of W49B has been estimated to be $\sim5,000$ yrs old, which supports most of the ejecta has been already heated \citep[e.g.,][]{2018A&A...615A.150Z}. In fact, the Fe-rich spectrum of W49B also supports that a large amount of the innermost ejecta has been heated. In this study, we thus assume that the observed Ti/Fe mass ratio reflects these elements produced through the Si burning during the SN explosion.

\subsection{Core-collapse supernova models}
In Figure \ref{fig:f2} (a), we compare the observed Ti/Fe ratio with those of the core-collapse supernova models in \cite{2001ApJ...555..880N} and \cite{2003ApJ...598.1163M}. We found almost all the core-collapse models are above our upper limit. In these models, the existence of the $\alpha$-rich freeze-out leads to the high Ti/Fe ratio. Therefore, the low Ti/Fe ratio provides strong constraints on the models. We have also checked the Ti/Fe ratio predicted by a series of recent core-collapse models constructed by \cite{2016ApJ...821...38S}, and found that Ti/Fe $\ge 1.9\times 10^{-3}$ in all the models. In addition, \cite{2023ApJ...948...80L} have performed a parameter study of jet-driven supernovae and their models, which have a large amount of $^{56}$Ni ($> 0.1 M_\odot$), also show a high Ti/Fe mass ratio of Ti/Fe $> 1.8\times 10^{-3}$.

The low Ti/Fe ratio we observed may mean a low explosion energy of its explosion, which could realize a weak $\alpha$-rich freeze-out. On the other hand, such a weak explosion is likely associated with a (relatively) low-mass star that can explode easily \citep[e.g.,][]{2016ApJ...818..124E,2016MNRAS.460..742M}. In that case, strong asymmetry during the shock stalling phase is not expected, thus a more-or-less spherical explosion with a weak neutron-star kick is an expected outcome \citep[e.g.,][]{2019PASJ...71...98N}. This does not follow the observational characteristics of the highly asymmetric remnant W49B.

In the previous studies of W49B, the abundances of Si and S were found to be lower than those in core-collapse models with high progenitor masses of $>15 M_{\odot}$ \citep[e.g.,][]{2018A&A...615A.150Z,2020ApJ...893...90S}. This stems from an increasing extent of Si/O layer (and thus increasing amount of Si and S) toward the more massive progenitors \citep[see the models in][]{1995ApJS..101..181W,2006NuPhA.777..424N,2016ApJ...821...38S,2018ApJ...860...93S}. Therefore, a progenitor with a large CO core, which is thought to be the origin of jet-driven supernovae, such as Gamma-Ray Bursts (GRBs) \citep[e.g.,][]{2007ApJ...654..938K,2013ApJ...764...50L}, is not favored for W49B.

In summary, the X-ray observations of W49B show that most of the current core-collapse supernova models have difficulties to explain the nature of W49B. Specifically, we have found that the Ti emissions, as expected from an asymmetric explosion of a massive star, including jet-driven explosions, are lacking with a tight upper limit on the amount of stable Ti produced at the explosion. The small amount of Ti could still be reconciled by a weak explosion of a low-mass star under the core-collapse SN scenario, but then the extremely asymmetric structure is not expected from such a model, and thus this interpretation is not favored either. 
 
\subsection{Type Ia supernova models}
The observed Ti/Fe ratio is also useful for constraining the models for SNe Ia. In particular, sub-$M_{\rm Ch}$ (sub-Chandrasekhar-mass) double-detonation models with a thick He shell, which is a variant within the sub-$M_{\rm Ch}$ model category, are expected to produce a large amount of stable Ti at the He-shell detonation \citep[e.g.,][]{2010A&A...514A..53F,2010ApJ...719.1067K,2020ApJ...888...80L}. Even without a thick He layer, e.g., violent WD merger models, the sub-$M_{\rm Ch}$ models generally predict a non-negligible amount of $^{44}$Ti ($\sim$10$^{-4}$--10$^{-3}~M_\odot$) produced through the QSE burning regime \citep[e.g.,][]{2022MNRAS.517.5260P,2022ApJ...932L..24R,2023MNRAS.519L..74K}. The small Ti/Fe ratio we observed can rule out some of these models.

In Figure \ref{fig:f2} (b) we have compared our Ti/Fe upper limit with those in theoretical calculations for SNe Ia, and found that the low-mass WD progenitors with $\lesssim$ 1.1--1.2 $M_\odot$ are above our upper-limit. Here, the sub-$M_{\rm Ch}$ SN Ia models have a tendency for a lower mass progenitor to produce a larger amount of Ti. This tendency can be explained by two reasons: (1) low-mass WDs require a thicker He layer to be exploded, and (2) SNe Ia from lower-mass WDs have a larger QSE layer relative to the NSE layer. 

One possible scenario to explain the Ti/Fe mass ratio in W49B is a sub-$M_{\rm Ch}$ SNe Ia model with $\gtrsim$ 1.1 $M_\odot$ with a thin He layer \citep[i.e., the D$^6$ model proposed in][]{2018ApJ...854...52S}. However, such double-detonation models usually show a symmetric Fe distribution, which may be difficult to explain the shape of W49B. In addition, the high Mn/Cr ratio reported in \cite{2018A&A...615A.150Z} indicates the difficulty of the sub-$M_{\rm Ch}$ SN Ia models, including the violent merger model, as the origin of W49B. In the case of the sub-$M_{\rm Ch}$ models, unnaturally high initial metallicity must be assumed to reproduce the high abundance of neutron-rich elements such as Mn \citep[e.g.,][]{2015ApJ...801L..31Y,2018A&A...615A.150Z,2020ApJ...888...80L}. 

Near-$M_{\rm Ch}$ SN Ia models would more naturally explain the observational properties of W49B than the sub-$M_{\rm Ch}$ models. We also show some near-$M_{\rm Ch}$ SN Ia models \citep[three-dimensional delayed-detonation models in][]{2013MNRAS.429.1156S} in Figure \ref{fig:f2} (b), where all of these fall within our Ti/Fe limit. In the case of near-$M_{\rm Ch}$ WDs, a large amount of stable Ti (=$^{50}$Ti) could be produced at the innermost region \citep[e.g.,][]{2018ApJ...861..143L,2021ApJ...913L..34O}. The innermost region of the near-$M_{\rm Ch}$ SNe Ia at $<$ 0.2 $M_\odot$ (i.e., at $\rho > 2 \times 10^8$ cm$^{-3}$ g) is consumed in the neutron-rich NSE regime \citep[e.g.,][]{1986A&A...158...17T,1999ApJS..125..439I}, where density-driven electron capture generates a neutron excess independent of the progenitor metallicity. While some near-$M_{\rm Ch}$ explosions with extremely high central-density may exceed our Ti/Fe limit due to this effect, we conclude that the near-$M_{\rm Ch}$ explosions are currently the best candidate to explain the observed Ti/Fe ratio, as seen in Figure \ref{fig:f2}. As for the asymmetric shape of W49B, asymmetric bouyant bubbles expected in the deflagration phase of near-$M_{\rm Ch}$ explosions \citep[e.g.,][]{2000astro.ph..8463K} may explain such a highly-asymmetric structure \citep{2010ApJ...712..624M}.

\section{Summary and Conclusions}

The origin of the W49B supernova remnant remains a topic of lively debate. One of the possible scenarios proposed so far is a jet-driven core-collapse supernova. This type of explosion is characterized by intense asymmetry, resulting in distinct regions of high entropy, where the $\alpha$-rich freezeout is expected to occur. This freezeout process is responsible for the production of elements through the capture of $\alpha$ particles, including elements such as $^{44}$Ti and $^{48}$Cr (=$^{48}$Ti). In this study, to shed light on the identity of the progenitor of the remnant, we have focused on the amount of stable Ti, specifically $^{48}$Ti. The presence and amount of $^{48}$Ti serve as valuable indicators of the occurrence of the high entropy process within the explosion.

From the {\it Suzaku} observations of the W49B remnant, we have found that there is no clear evidence for the Ti line and set the upper limit of $M_{\rm Ti}/M_{\rm Fe} < 8.2 \times$ 10$^{-4}$, with a confidence level of 99\%. Remarkably, this limit has useful implications; it effectively rules out a significant fraction of core-collapse supernovae and Type Ia supernova scenarios. A highly-asymmetric core-collapse SN including a jet-driven explosion, which has been suggested as the origin of W49B \citep[e.g.,][]{2007ApJ...654..938K,2013ApJ...764...50L}, is not favored, even if using the upper limit of $M_{\rm Ti}/M_{\rm Fe} < 1.9 \times$ 10$^{-3}$ with SPEX. The remaining/favored possibility is an SN Ia originating from a relatively massive white dwarf ($\gtrsim 1.1-1.2 M_\odot$), either through sub-$M_{\rm Ch}$ double-detonation or near-$M_{\rm Ch}$ delayed-detonation scenario. We have argued that the near-$M_{\rm Ch}$ scenario is more likey the case, once the abundances of neutron-rich elements and the morphology of Fe emission are combined with the present results. X-ray calorimeter missions, such as XRISM \citep{2018SPIE10699E..22T}, have the potential to provide even more precise and restrictive constraints on the nature of the progenitor.

\acknowledgments
This study was done before the XRISM observation of W49B as already reported in the ASJ spring meeting in 2023 (the presentation id is N17a). This work was supported by the Japan Society for the Promotion of Science (JSPS) KAKENHI grant No. JP19K14739, JP23K13128 and JP20H00174.

\bibliography{sample63}{}
\bibliographystyle{aasjournal}



\end{document}